\documentclass[twoside,twocolumn,9pt]{article}
\usepackage{comment}
\usepackage{extsizes}
\usepackage[super,sort&compress,comma]{natbib} 
\usepackage[version=3]{mhchem}
\usepackage[left=1.5cm, right=1.5cm, top=1.785cm, bottom=2.0cm]{geometry}
\usepackage{balance}
\usepackage{mathptmx}
\usepackage{sectsty}
\usepackage{graphicx} 
\usepackage{lastpage}
\usepackage[format=plain,justification=justified,singlelinecheck=false,font={stretch=1.125,small,sf},labelfont=bf,labelsep=space]{caption}
\usepackage{float}
\usepackage{fancyhdr}
\usepackage{fnpos}
\usepackage[english]{babel}
\addto{\captionsenglish}{%
  \renewcommand{\refname}{Notes and references}
}
\usepackage{array}
\usepackage{droidsans}
\usepackage{charter}
\usepackage[T1]{fontenc}
\usepackage[usenames,dvipsnames]{xcolor}
\usepackage{setspace}
\usepackage[compact]{titlesec}
\usepackage{hyperref}

\usepackage{epstopdf}

\definecolor{cream}{RGB}{222,217,201}

\begin{document}

\pagestyle{fancy}
\thispagestyle{plain}
\fancypagestyle{plain}{

\renewcommand{\headrulewidth}{0pt}
}

\makeFNbottom
\makeatletter
\renewcommand\LARGE{\@setfontsize\LARGE{15pt}{17}}
\renewcommand\Large{\@setfontsize\Large{12pt}{14}}
\renewcommand\large{\@setfontsize\large{10pt}{12}}
\renewcommand\footnotesize{\@setfontsize\footnotesize{7pt}{10}}
\makeatother

\renewcommand{\thefootnote}{\fnsymbol{footnote}}
\renewcommand\footnoterule{\vspace*{1pt}% 
\color{cream}\hrule width 3.5in height 0.4pt \color{black}\vspace*{5pt}} 
\setcounter{secnumdepth}{5}

\makeatletter 
\renewcommand\@biblabel[1]{#1}            
\renewcommand\@makefntext[1]% 
{\noindent\makebox[0pt][r]{\@thefnmark\,}#1}
\makeatother 
\renewcommand{\figurename}{\small{Fig.}~}
\sectionfont{\sffamily\Large}
\subsectionfont{\normalsize}
\subsubsectionfont{\bf}
\setstretch{1.125} %In particular, please do not alter this line.
\setlength{\skip\footins}{0.8cm}
\setlength{\footnotesep}{0.25cm}
\setlength{\jot}{10pt}
\titlespacing*{\section}{0pt}{4pt}{4pt}
\titlespacing*{\subsection}{0pt}{15pt}{1pt}

\fancyfoot{}
\fancyfoot[LO,RE]{\vspace{-7.1pt}\includegraphics[height=9pt]{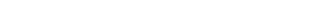}}
\fancyfoot[CO]{\vspace{-7.1pt}\hspace{13.2cm}\includegraphics{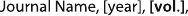}}
\fancyfoot[CE]{\vspace{-7.2pt}\hspace{-14.2cm}\includegraphics{head_foot/RF}}
\fancyfoot[RO]{\footnotesize{\sffamily{1--\pageref{LastPage} ~\textbar  \hspace{2pt}\thepage}}}
\fancyfoot[LE]{\footnotesize{\sffamily{\thepage~\textbar\hspace{3.45cm} 1--\pageref{LastPage}}}}
\fancyhead{}
\renewcommand{\headrulewidth}{0pt} 
\renewcommand{\footrulewidth}{0pt}
\setlength{\arrayrulewidth}{1pt}
\setlength{\columnsep}{6.5mm}
\setlength\bibsep{1pt}

\makeatletter 
\newlength{\figrulesep} 
\setlength{\figrulesep}{0.5\textfloatsep} 

\newcommand{\topfigrule}{\vspace*{-1pt}% 
\noindent{\color{cream}\rule[-\figrulesep]{\columnwidth}{1.5pt}} }

\newcommand{\botfigrule}{\vspace*{-2pt}% 
\noindent{\color{cream}\rule[\figrulesep]{\columnwidth}{1.5pt}} }

\newcommand{\dblfigrule}{\vspace*{-1pt}% 
\noindent{\color{cream}\rule[-\figrulesep]{\textwidth}{1.5pt}} }

\makeatother

\twocolumn[
  \begin{@twocolumnfalse}
{\includegraphics[height=30pt]{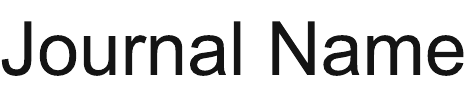}\hfill\raisebox{0pt}[0pt][0pt]{\includegraphics[height=55pt]{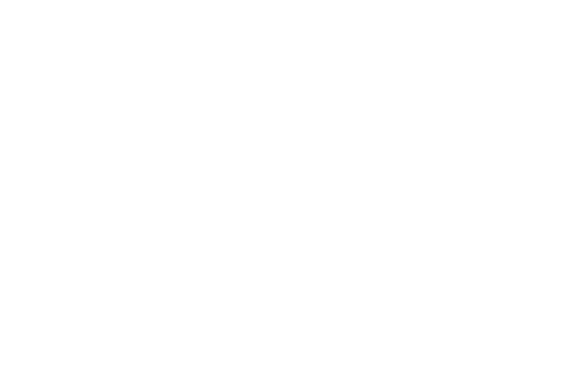}}\\[1ex]
\includegraphics[width=18.5cm]{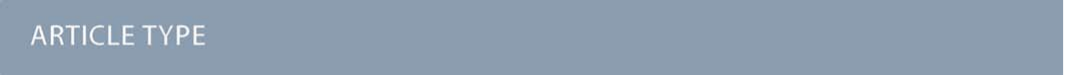}}\par
\vspace{1em}
\sffamily
\begin{tabular}{m{4.5cm} p{13.5cm} }

\includegraphics{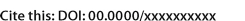} & \noindent\LARGE{\textbf{An active soft glassy rheology model }} \\
\vspace{0.3cm} & \vspace{0.3cm} \\

 & \noindent\large{Raffaele Mendozza$^{\ast}$\textit{$^{a}$}, Tobias Müller\textit{$^{a}$}, Peter Sollich\textit{$^{a,b}$} } \\

\includegraphics{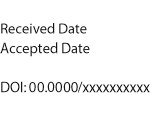} & \noindent\normalsize{Biological materials such as the cytoskeleton and confluent cell monolayers are active, dense systems continuously subjected to internal stresses and strains, making their rheological characterization essential. While activity in soft matter can be modeled across multiple length scales, its mechanical consequences remain strongly model dependent and no unified theoretical framework has yet emerged. Here, we study the rheology of dense active amorphous materials using the Soft Glassy Rheology (SGR) model, incorporating activity at the mesoscopic scale of local elements as a stochastic strain rate that is persistent on some timescale $\tau_p$. We show that activity opens a long-time relaxation channel, driving a crossover from SGR-like power-law rheology to Maxwell-like behavior at the lowest frequencies. Combining analytical arguments in limiting regimes with numerical simulations, we characterize the resulting fluidisation time scale and its dependence on the activity parameters, which shows strong analogies with effective temperatures introduced elsewhere  
that similarly encode activity-induced fluidisation. Our active SGR model provides a minimal mesoscopic route to understanding how driving by activity modifies the rheology of dense amorphous materials.
}\\

\end{tabular}

 \end{@twocolumnfalse} \vspace{0.6cm}

  ]

\renewcommand*\rmdefault{bch}\normalfont\upshape
\rmfamily
\section*{}
\vspace{-1cm}

\footnotetext{\textit{$^{a}$~University of Göttingen, Institute for Theoretical Physics, 37077 Göttingen, Germany; E-mail: peter.sollich@uni-goettingen.de}}
\footnotetext{\textit{$^{b}$~King's College London, Department of Mathematics, Strand, London WC2R 2LS, U.K. }}

\section{Introduction}
\label{sec:intro}
Active matter broadly denotes materials that convert internal or environmental energy into mechanical work and are therefore intrinsically out of equilibrium, ranging from engineered Janus colloids to biological systems across many length and time scales \cite{Gompper_2020, Ramaswamy_2017}. While much of active matter physics has focused on dilute suspensions of self-propelled particles, many biological systems are instead crowded, dense, and often close to glassy or jammed states \cite{Sadhukhan_2024}. Understanding how activity modifies the relaxation and mechanical response of such crowded materials is therefore central, especially for cells and subcellular assemblies whose biological functions are tightly linked to their mechanical properties \cite{Janmey_2004}. This problem has been addressed extensively through particle-based simulations of active glasses and dense active suspensions \cite{Loi_2008,Berthier_2014,Mandal_2016,Flenner_2016,Bechinger_2016,Fernandez_Barrat_Henkes_2017}), and via more theoretical approaches based on e.g.\ active mode-coupling theory and random first-order transition theory \cite{Nandi_2017,Nandi_2018}. A complementary route is to construct mesoscopic descriptions, such as elastoplastic models, in which relaxation proceeds by local yield events and the associated elastic stress redistribution \cite{Fielding_2023,Barton_2017,Nicolas_2018,Krajnc_2018,Lin_2023,Ghosh_2025}. These approaches show that activity can compete with local yielding and, in suitable regimes, induce fluidisation, although the specific mechanism remains model dependent and a unified understanding is still lacking. Here, we study activity within the Soft Glassy Rheology model \cite{Sollich_1997}, a minimal coarse-grained framework for the mechanical response of dense amorphous materials under shear \cite{Nicolas_2018}. This has previously also been used as a phenomenological description of biological soft matter such as the cytoskeleton, where power-law rheology is observed \cite{Fabry_2001,Fabry_2003,Mandadapu_2008}, though without explicit inclusion of active processes. As proposed by Sollich and Cates \cite{Sollich_2012}, we slightly modify the model by introducing an upper bound to the yielding rate, corresponding to the fastest (passive) relaxation channel. We then incorporate activity as a local strain-rate drive specified by a typical strain rate $v_0$ and persistence time $\tau_p$, and compute the complex shear modulus. We show that activity produces a long-time fluidisation of the material, characterized by a crossover to a Maxwell-like response at low frequencies, in qualitative agreement with active elastoplastic descriptions \cite{Barton_2017,Krajnc_2018,Ghosh_2025,Lin_2023}. This suggests a generic activity-induced fluidisation mechanism beyond the details of the chosen mesoscopic model.

In Sec.~\ref{sec:Models_Methods} we start with a brief introduction to the SGR model and detail the specific form of yield rate that we use. We proceed with the introduction of activity into the model, and describe the theoretical approach used to compute the linear viscoelastic spectrum from the model. Then, in Sec.~\ref{sec:Results_Siscussion} we study the effects of activity on the spectrum, combining analytical and numerical approaches. Sec.~\ref{sec:Discussion_Conclusions}, finally, contains a summary and discussion of our results.

\section{Models and Methods}
\label{sec:Models_Methods}
\subsection{The SGR model with capped yield rate}
\label{subsec:SGR_model}
\begin{figure}[htb]
    \includegraphics{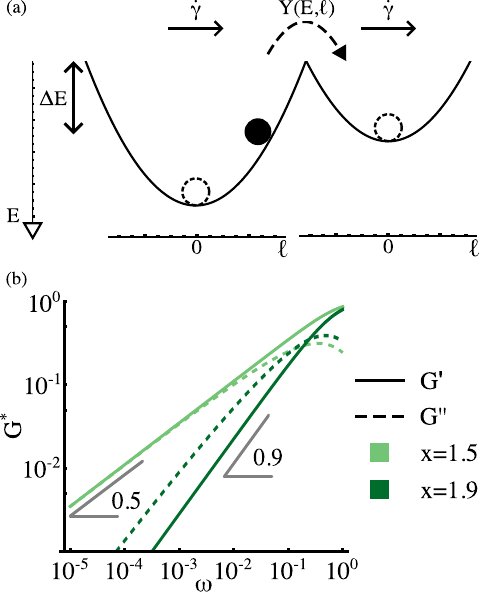}
    \caption{\label{fig:sketch}
    (a) Sketch of element dynamics in the SGR model. The elastic energy of a local element is quadratic in the local strain $\ell$. The latter changes in line with the externally applied shear rate $\dot\gamma$. Additionally, a yield event takes place when the local elastic energy becomes of the order of the  yield energy $E$, with rate $Y(E,\ell)$. After a yield, a new value of $E$ is assigned and $\ell$ is reset to zero. (b) Example viscoelastic spectra of the passive SGR model, for two different values of $x<2$. For low enough frequencies, the complex shear modulus follows $G^*(\omega) \sim \left(i\omega\right)^{x-1}$, so both the elastic modulus
    $G'={\rm Re}(G^*)$ and the viscous modulus $G''={\rm Im}(G^*)$ vary with the same power law. Both figures are adapted from Ref.~\cite{Sollich_1998}.
    }
\end{figure}

The soft glassy rheology (SGR)~\cite{Sollich_1997,%
Sollich_1998} model was developed to provide a mesoscopic description for the viscoelastic spectrum, and more general nonlinear shear rheology, of soft amorphous materials. It incorporates their main features of structural disorder and metastability via a distribution of yield stresses, in what can be viewed as a mean-field approximation to elastoplastic modelling approaches \cite{Nicolas_2018}. 

One of the main motivations for the development of the SGR model was the ubiquitous power-law rheology exhibited by many complex fluids, which suggests that a common mechanism might be at play. This is reflected in the relatively generic assumptions of the model, which we summarize next as they form the basis for the active SGR model we propose below. Each mesoscopic portion of the material (called ``element''
) is described by two degrees of freedom: the local strain $\ell$ relative to some local equilibrium configuration, and a yield energy $E$. The local elastic energy is taken as quadratic in $\ell$, with the same stiffness $k$ for all elements. Yielding, i.e.\ a plastic rearrangement that results in a new local equilibrium configuration, takes place with a  state-dependent yield rate $Y(E,\ell)$ that increases as the elastic energy $k\ell^2/2$ approaches the yield energy $E$. 
After a yield, the element is assigned a new yield energy $E$ drawn from an {\em a priori} distribution $\rho(E)$, which can be viewed as a density of states. The local strain is reset to $\ell=0$ at the same time.
Finally, when external shear strain is applied at rate $\dot\gamma$, the SGR model assumes that all elements increase their local strain $\ell$ in line with this external strain. Combining all assumptions, one can write for the time evolution of the distribution $P(E,\ell;t)$ of elements across the different values of $E,\ell$ the
following master equation:
    \begin{equation}
        \label{eqn:SGR_master_equation}
         \partial_t P(E,\ell;t) = - \dot{\gamma} \partial_\ell P - Y(E,\ell) P + \langle Y(E,\ell) \rangle
         \rho(E)\delta(\ell) ,
    \end{equation}
where $\langle \cdot \rangle$ denotes the average over $P$ itself. The first term in~(\ref{eqn:SGR_master_equation}) describes the increase of local strains with the external strain, while the second and third term represent the yield events, with  the Dirac delta function accounting for the reset of the local strain to $\ell=0$ after a yield event. See Fig.~\ref{fig:sketch} (a) for a sketch summarising the dynamical process that the SGR model describes. To characterize the rheological response, finally, the SGR model assumes that the macroscopic stress is the average of the local stresses of all elements, i.e.\ $\sigma=k\langle \ell\rangle$.

We will use for the yield rate the form
    \begin{equation}
        \label{eqn:yield rate}
        Y(E,\ell) = \Gamma_0 \min\left\{1 , \exp\left[- \left( E - \frac{1}{2} k \ell^2\right)/x\right]\right\} .
    \end{equation}
Here the exponential dependence describes activated yielding governed by the energy barrier to yield, $E-k\ell^2/2$, with an effective temperature $x$ that encodes mechanical noise arising from yield events elsewhere in the material. In the original SGR model this exponential behaviour was assumed to apply even when this energy barrier formally becomes negative, where an activation picture becomes hard to defend physically. The ``min'' in~(\ref{eqn:yield rate}) therefore adds a physical cap on the yield rate, so that the yield rate cannot exceed the underlying attempt frequency $\Gamma_0$ even once the strain $\ell$ exceeds the \textit{yield strain} $\ell_y=\sqrt{2E/k}$. This modification was in fact already suggested in Ref.~\cite{Sollich_2012,Rouyer_2008} to make the SGR model more physically realistic. 
To summarize it in words, after straining, elements with low $E$ -- which are \textit{shallow} traps in the sense of the Bouchaud trap model \cite{Monthus_1996} -- all have the same constant yield rate, while deep traps (elements with comparatively large $E$) retain the activated form of their yielding dynamics. For the linear rheology of the original ``passive'' SGR model, the capping of the yield rate makes no difference, as all $\ell$ are then infinitesimal and all traps are effectively ``deep''.
As a result, even with the capping one finds in the fluid regime of the model ($x>1$) the expression for the viscoelastic spectrum derived for the original SGR model~\cite{Sollich_1997}, 
    \begin{equation}
        G^*(\omega) = \left\langle \frac{i \omega \tau}{ 1 + i \omega \tau} \right\rangle,
    \end{equation}
    This is a weighted superposition of Maxwell modes with relaxation time $\tau = e^{E/x}$. We will recall the derivation of this result in Sec.~\ref{subsec:Linear_expansion_approach} below. 
    For an exponential distribution of yield energies as chosen above, the shear modulus scales asymptotically as $G^*(\omega) \sim (i \omega)^{x-1}$ for $\omega \ll 1$, as shown in Fig.~\ref{fig:sketch}(b). This scaling applies for $1<x<2$, which is the regime with the clearest ``glassy'' effects and therefore our focus also in this paper.

For simplicity, we will in the following fix time units so that $\Gamma_0 =1 $, and energy units so that typical values of $E$ are also of order unity. Specifically, we will take $\rho(E)$ as exponential, $\rho(E)=e^{-E}$ for $E>0$. Finally, we will assume that strains have been scaled so that also the local elastic constant $k$ is unity, implying that $\ell$ is measured in units of a typical local yield strain. 

In closing this section we note that, in contrast to the case of the passive SGR model described so far, the yield rate capping will matter in the active SGR model that we turn to next: there, the activity will generically produce non-negligible local strains that can become of the order of the local yield strain.

\subsection{Active SGR model}
\label{subsec:Active_SGR}
In constructing an active version of the SGR model that could capture the rheology of active systems with glassy features \cite{Janssen_2019}, we want to include activity as a local injection of energy that breaks detailed balance (which in the original SGR model and variants with other choices for the yield rate is satisfied \cite{Sollich_2012}). How to do this most realistically is a subject of ongoing research \cite{ Gompper_2020,Ghosh_2025,Lin_2023,Woillez_2020} .
Here, inspired by biological systems such as biopolymer networks driven by molecular motors or active confluent cell monolayers, we choose to introduce activity as a {\em local strain rate} \cite{Fodor_2014,Fodor_2015,Ahmed_2015} . Specifically, 
we assume that each mesoscopic element of the system is subject to an independent, local active strain rate $v$. The latter is taken as a zero mean stochastic process with some variance $v_0^2$ and a correlation time $\tau_p$, for which we take as the simplest option Ornstein-Uhlenbeck (OU) process: 
    \begin{equation}
        \label{eqn:OU_activity}
        \dot{v} = - \frac{1}{\tau_p} v + \sqrt{\frac{2 v_0^2}{\tau_p}} \xi,
    \end{equation}
    where $\xi(t)$ is Gaussian white noise of unit variance.
With this approach, activity enters the master equation as an additional drift term in the time evolution of $\ell$, making the local strain rate (between two yield events) $\dot\gamma + v$ rather than $\dot\gamma$ as before.
With these assumptions, the active SGR model can be described by a master equation for the extended distribution $P(E,\ell,v;t)$:
    \begin{equation}
    \label{eqn:P_master_equation_with_activity}
        \begin{aligned}
                \partial_t P(E,l,v;t) = &- (\dot\gamma+v)\partial_\ell\, P - Y(E,\ell)P + \langle Y(E,\ell) \rangle \rho(E) \delta(\ell) \\
                &+\frac{1}{\tau_p}\partial_v(vP) +\frac{v_0^2}{\tau_p}\partial_v^2 P .
        \end{aligned}
    \end{equation}
Here the second line encodes the OU dynamics of the local active strain rate $v$, which couples to the time evolution of the local strain $\ell$ via $(\dot\gamma+v)$ in the first line.
Note that, conversely, the dynamics of $\ell$ or $E$ does not affect the time evolution of the local active strain rate $v$. This is consistent with the underlying physical picture that activity is induced in our setting by some ``additional'' agent (e.g.\ motors moving on a passive network) that is not itself affected by the yield and deformation of the material. In particular, we assume that $v$ is not subject to any resetting at yield, and
accordingly the average $\langle Y(E,l)\rangle$ in~(\ref{eqn:P_master_equation_with_activity}) is to be read as $\int dE\,d\ell\,Y(E,l)P(E,l,v)$ without integration over $v$. By contrast in the calculation of physical observables one does, of course, have to average also over $v$.
We note finally that the fact that the dynamics of $E$ and $\ell$ does not affect the one of $v$ can also be seen mathematically by integrating eq.~(\ref{eqn:P_master_equation_with_activity}) over $E$ and $\ell$, which gives for the time evolution of the marginal distribution $P(v;t)$ the standard Fokker-Planck equation for an OU process,
\begin{equation}
\label{eq:dPv_dt}                        \partial_t P(v;t) = \frac{1}{\tau_p}\partial_v(vP) +\frac{v_0^2}{\tau_p}\partial_v^2 P .
\end{equation}
For large $t$, $P(v;t)$ therefore approaches the expected Gaussian steady state distribution 
\begin{equation}
P_{\rm ss}(v) = (2\pi v_0^2)^{-1/2}e^{-v^2/(2v_0^2)}\ .
\label{eq:Pv_ss}
\end{equation}
    
%\end{itemize}
\subsection{Linear viscoelastic spectrum}
\label{subsec:Linear_expansion_approach}
We will focus in this paper on a study of the linear viscoelastic spectrum of the active SGR model, i.e.\ the complex shear modulus $G^*(\omega)$.
We consider materials close to but above their glass transition, $x>1$, which ensures that the system reaches a (non-equilibrium) stationary state for large enough $t$. When subject to an oscillatory external strain $\gamma(t)={\rm Re}(\gamma_0 e^{i\omega t})$ of infinitesimal amplitude $\gamma_0$ and frequency $\omega$, the time-dependent stress after any transients is then $\sigma(t) = \gamma_0{\rm Re}[G^*(\omega)e^{i\omega t}]$ to leading order; the leading correction terms would be $O(\gamma_0^3)$ because of the symmetry of the stress response ($\sigma\to -\sigma$ for $\gamma_0\to -\gamma_0$). 

To calculate $G^*(\omega)$, one has to find the linear response of the distribution $P(E,\ell,v;t)$ to the applied strain, which -- again after any transients -- has the form 
\begin{equation}
P(E,\ell,v;t) = P_{\rm ss}(E,\ell,v) + \gamma_0 {\rm Re}[P^{(1)}(E,\ell,v;\omega)e^{i\omega t}] .
\label{eq:linear_response_P}
\end{equation}
Here $P_{\rm ss}$ is the steady state distribution. Along with the linear response $P^{(1)}$,
this can in principle be calculated from eq.~(\ref{eqn:P_master_equation_with_activity}). Averaging gives the macroscopic stress $\sigma=\langle \ell\rangle$, so that the complex shear modulus can be read off as $G^*(\omega) = 
\int dE\,d\ell\,dv\,\ell\, P^{(1)}(E,\ell,v;\omega)$. The stationary state $P_{\rm ss}$ makes no contribution to the stress as usual; mathematically this can be seen from the symmetry of the master equation under $(\ell , v ) \to (- \ell, -v)$.

While in the passive SGR model the above quantities can be calculated relatively easily, even $P_{\rm ss}$ cannot generally be determined in closed form in the active SGR model, because of the coupling between $l$ and $v$. Analytical progress can be made, however, in two limiting cases where timescale separation can be exploited: extremely persistent activity ($\tau_p\to\infty$) and activity with vanishing temporal correlations ($\tau_p\to 0$). We will refer to these two regimes as, respectively, the \textit{adiabatic} and \textit{Brownian} limits. 

\section{Results and discussion}
\label{sec:Results_Siscussion}
\subsection{Adiabatic limit}
\label{subsec:adiabatic_limit}
\begin{figure}[htb]
    \centering
    \includegraphics{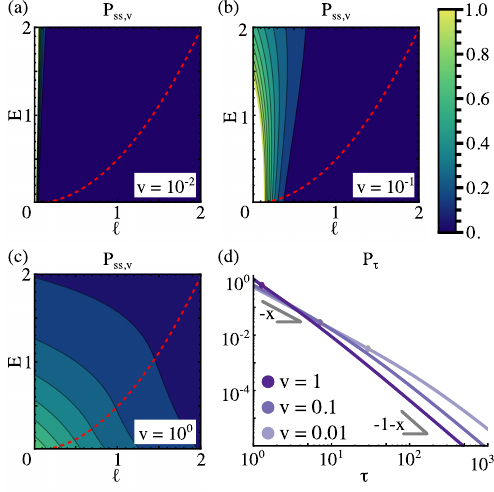}
    \caption{ (a-c) Steady state probability distributions $P_{\rm ss}(E,\ell|v)$ of $E$ for $x=1.5$ and different fixed values of $v$ as shown. The dashed red line is the yield strain line $\ell = \sqrt{2 E}$, where the yield rate reaches its maximum value $\Gamma_0$. For small $v$ the distribution approaches the original SGR result $P_{\rm ss}(E,\ell)\propto \rho(E)\tau(E)\delta(\ell)$ with $\tau(E)=e^{E/x}$, which is concentrated around small strains $\ell$. As $v$ 
    increases, the distribution shifts towards larger $\ell$. (d) Steady state distribution of trap lifetimes $\tau$, plotted in log-log scale. For fixed $v$, the distribution crosses over from a power law with exponent $-x$ at small $\tau$, to one with exponent $-1-x$ at large $\tau$. This shows that $v$ introduces a timescale distinguishing between shallow (small $E$ and $\tau$) and deep (large $E$ and $\tau$) traps. The coloured dots indicate the crossover time $\tau^*$, discussed in detail in Sec.~\ref{subsec:adiabatic_limit}.
    }
    \label{fig:ss_adiabatic_limit}
\end{figure}

\noindent 
In the limit of extremely persistent activity,
the dynamics of $v$ is very slow compared to that of $E$ and $\ell$. We can therefore solve for $P_{\rm ss}(E,\ell|v)$ and $P^{(1)}(E,\ell|v;\omega)$ first for fixed $v$, and then average over the steady state distribution~(\ref{eq:Pv_ss}) of $v$. To see this explicitly, one can insert into eq.~(\ref{eqn:P_master_equation_with_activity}) the decomposition $P(E,\ell,v,t)=P(E,\ell;t|v)P_{\rm ss}(v)$ and take the limit $\tau_p\to \infty$. All contributions from the second line then disappear and one obtains for $P(E,\ell;t|v)$ the time evolution equation
   \begin{equation}
                \partial_t P(E,\ell;t|v) = - (\dot\gamma+v)\partial_\ell\, P - Y(E,\ell)P +\langle Y(E,\ell) \rangle \rho(E) \delta(\ell)\,.
                \label{eqn:P_cond}
    \end{equation}
For each fixed $v$, we are then solving effectively a problem of ``superposition rheology'' \cite{Dhont_2001,Lee_2024}, where -- as described by the first term on the r.h.s.\ of eq.~(\ref{eqn:P_cond})
-- one studies the linear stress response to an external strain $\gamma(t)$ that is superimposed on steady shear with rate $v$. 
The unperturbed steady state can be taken from the known steady shear solution~\cite{Sollich_1998}, and reads for $v>0$
\begin{equation}
P_{\rm ss}(E,\ell|v) = \frac{Y_v}{v} \,\exp{\left \{-\frac{1}{v} \int_0^\ell d\ell' \, Y(E,\ell ')\right\}} \rho(E) \Theta(\ell) .
\end{equation}
Here the Heaviside theta function $\Theta(\ell)$ arises from the assumption $v>0$, while $Y_v$ is the total yield rate, i.e.\ the average of $Y(E,l)$ over $P_{\rm ss}(E,\ell|v)$. As illustrated by the example plots in Fig.~\ref{fig:ss_adiabatic_limit}, 
this expression interpolates between a passive, unsheared soft glassy material for $v\to 0$ and a Maxwell material with a single relaxation time $1/\Gamma_0$ for large $v$. To see this, consider first the small $v$ limit, where the exponent becomes large and negative already for small $\ell$. One can then approximate $Y(E, \ell) \approx Y(E,0) = 1/ \tau(E)$ to obtain $P_{\rm ss} (E, \ell|v) =(Y_v/v)\rho(E) e^{-{\ell}/[{v \tau(E)}]}$. Writing the prefactor as $Y_v\tau(E)/[v\tau(E)]$ shows that for small $v$ the distribution approaches 
$P_{\rm ss} (E, \ell) \propto \rho(E)\tau(E)\delta(\ell)$, which is the passive SGR result.
In the opposite limit of large $v$, typical strains in $P_{\rm ss}$ will become large so that the yield rate reaches its maximum $Y(E,\ell)\approx \Gamma_0$, giving explicitly
$P_{\rm ss} = \frac{\Gamma_0}{v} \rho(E) e^{- \Gamma_0 \ell / v}$ (with $\Gamma_0=1$ in our units). 

From the two limits above, it is easy to obtain the corresponding distributions of trap lifetimes $\tau$, which exhibit power law behaviour in both cases: $\sim \tau^{-x}$ for small $v$ and $\sim \tau^{-1-x}$ for large $v$.
For fixed, finite $v$ the distribution of $\tau$ shows a crossover between these two scalings, as illustrated in Fig.~\ref{fig:ss_adiabatic_limit}(d).
The intuition is that shallow traps, i.e.\ elements with small $E$ and $\tau$, yield sufficiently rapidly due to the mechanical noise that they are insensitive to the applied steady shear. Deep traps, on the other hand, acquire strains above their yield strain before they have time to yield by mechanical activation. The crossover time, i.e.\ the value of $\tau$ where $P_{\rm ss}(\tau|v)$ changes between the two power law scalings, will be discussed in the following Sec.~\ref{subsec:adiabatic_limit}. Qualitatively, however, we can anticipate already that the macroscopic stress response will involve both SGR-like and Maxwell-like features in the appropriate frequency ranges. 

Having established the steady state for fixed $v$, we can now turn to the response to small oscillatory strain. Inserting the analogue for fixed $v$ of the ansatz~(\ref{eq:linear_response_P}) into eq.~(\ref{eq:dPv_dt}) gives, from the terms of $O(\gamma_0)$, the relation
    \begin{equation}
        \label{eqn:adiabatic_Fourier_kernel}
        \left(i\omega-\mathcal L_v\right) P^{(1)}(E,l;\omega|v)
    = 
-i\omega \partial_\ell P_{{\rm ss}}(E,\ell|v) ,
    \end{equation}
where the linear operator $\mathcal{L}_v$ is defined as $\mathcal L_v f = -v\partial_\ell f -Y(E,\ell)f + \rho(E)\delta(\ell) \int dE'd\ell'\, Y(E',\ell')f(E',\ell') $. From eq.~(\ref{eqn:adiabatic_Fourier_kernel}) we can determine $P^{(1)}$ using a suitable Green's function, and from there find
the complex shear modulus for fixed $v$,
$G_v^*(\omega) = \int dE\,d\ell \, \ell P^{(1)}(E,\ell;\omega|v)$. 
The explicit expression for the latter can be cast, after some algebra, into the relatively simple form
    \begin{equation}
        \label{eqn:adiabatic_shear_mod_In_Jn_integrals}
        G^*_v(\omega)=-i\frac{\omega}{v}\left(\frac{I_1}{I_0}J_0-J_1\right) ,
    \end{equation}
    where 
    \begin{equation}
    I_n = \langle \ell^n e^{-i(\omega/v)\ell } \rangle_{{\rm ss}}, \qquad 
    J_n = \langle \ell^n y(\ell,E,\omega/v) \rangle_{{\rm ss}} ,
    \label{eq:I_J}
    \end{equation}
and
we have abbreviated $y(E,\ell,z) = \int_0^\ell d\ell' \, \frac{Y(\ell',E)}{v} e^{-i z (\ell - \ell')}$. The averages in eq.~(\ref{eq:I_J}) are over $P_{\rm ss}(E,\ell|v)$.

The overall complex shear modulus is finally obtained by averaging over the steady state distribution~(\ref{eq:Pv_ss}) of the OU process of $v$:
   \begin{equation}
   \label{eqn:adiabatic_shear_modulus}
       G^*(\omega) = %\displaystyle 
       \int
       dv \,
       P_{\rm ss}(v)
       G_v^*(\omega) .
   \end{equation} 
The numerical calculation of the integrals over $E$, $\ell$ and $v$ that arise in evaluating eq.~(\ref{eqn:adiabatic_shear_modulus}) requires some care because of the oscillatory nature of the integrands. In our implementation this limits the regime where we can obtain accurate results 
to $v_0^2 > 10^{-8}$ and $\omega > 10^{-3}$.

\begin{figure*}[htb]
    \centering \includegraphics{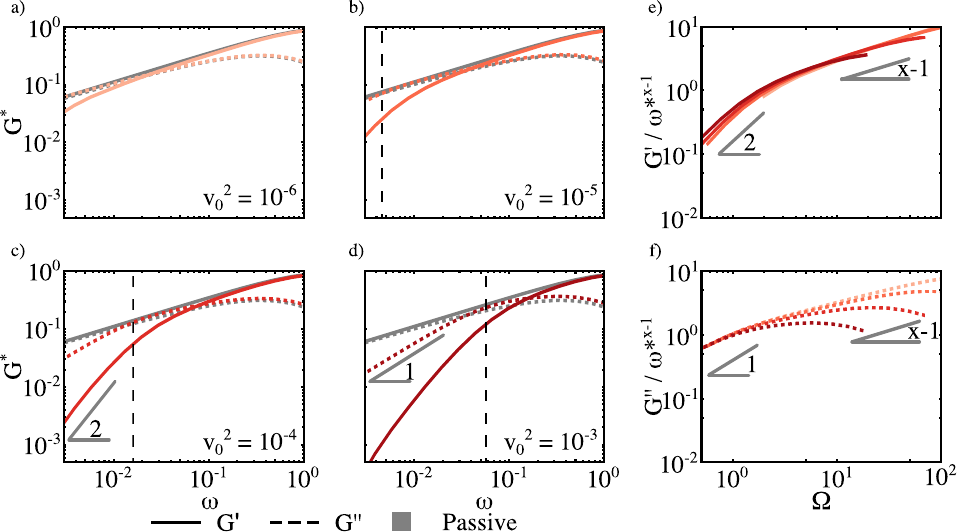}
    \caption{
    (a-d) Complex shear modulus $G^*$ in the adiabatic limit for $x= 1.5$, obtained for increasing variance of activity $v_0^2$, as reported in the figures. Activity induces a crossover at some frequency $\omega^*$, as discussed in the main text. The predicted values of $\omega^*$ are highlighted by the vertical black dashed lines. At frequencies below $\omega^*$, the viscoelastic spectrum crosses over to Maxwell behaviour as indicated by power law exponents of 2 and 1, respectively, for the elastic modulus $G'$ and the viscous modulus $G''$. For higher $\omega$, passive SGR behaviour is approached (gray lines), where $G'$ and $G''$ both scale as $\omega^{x-1}$. 
    (e,f) Rescaled viscoelastic spectra vs rescaled frequency $\Omega = \omega / \omega^*$. The curves are color-coded as in (a-d) and are consistent with collapse to a master curve in the limit of small $\omega^*$, where the crossover frequency becomes well separated from the upper end $\omega \sim 1$ of the frequency range of SGR behaviour.
    }
    \label{fig:spectrum_adiabatic_limit}
\end{figure*}

We show exemplary results from numerical integration of eq.~\ref{eqn:adiabatic_shear_modulus} for different $v_0^2$ in Fig.~\ref{fig:spectrum_adiabatic_limit}(a-d).
The computed spectra show Maxwell-like scaling at low frequency, and cross over to the passive SGR power law scaling $G^*\sim \omega^{x-1}$ for higher frequencies (while always keeping $\omega < 1$ to remain in the frequency window where effects not captured by SGR, from e.g.\ solvent viscosity, might start to matter). To estimate the crossover frequency, we propose a simple physical argument.
As discussed in Sec.~\ref{subsec:adiabatic_limit}, a fixed active strain rate $v\sim v_0$ introduces a timescale $\tau^*$ that distinguishes whether a local element will relax its stress via activation by mechanical noise, or via active strain that pushes it beyond its yield strain.
We expect the crossover to take place where the two processes take similar amounts of time. For activation this time is $\tau(E)=e^{E/x}$; for yielding by active shear it is $\sqrt{2E}/v_0$ because the yield strain $\sqrt{2E}$ has to be reached by strain with rate $\sim v_0$. Equating these two estimates gives a transcendental equation for $E$, which can be formally solved using Lambert functions.
Expressing the result in terms of $\tau(E)$, we find to leading order in logarithmic corrections the solution
\begin{equation}
\tau^* = \sqrt{\frac{x}{v_0^2}\ln\left( 2 x / v_0^2 \right)} .
\label{eq:taustar_adiabatic}
\end{equation}
The associated crossover frequency can then be estimated as $\omega^* = {2 \pi}/{\tau^*}$. This estimate is shown by the black dashed lines in Fig.~\ref{fig:spectrum_adiabatic_limit} and gives a good account of the location of the crossover between the two power law regions of $G^*(\omega)$ that we see numerically. Going further, one expects that for small enough $\omega^*$, where this crossover frequency is far below the upper limit $\omega\sim 1$ of the SGR power law regime, results for different $\omega^*$ should  scale onto a master curve. 
To check this, we introduce the rescaled frequency $\Omega = \omega / \omega^*$ and plot against this the rescaled modulus $G^*/{\omega^*}^{x-1}$. As shown in Fig.~\ref{fig:spectrum_adiabatic_limit}(e,f), our data are consistent with the expected approach to a master curve in the limit of small $\omega^*$. Furthermore, while the frequency range we can probe is limited by numerical accuracy, the master curves for $G'$ and $G''$ show behaviour consistent with a Maxwell viscoelastic spectrum for $\Omega < 1$ and with SGR scaling for $\Omega > 1$, confirming
the expected crossover.

\subsection{Brownian limit}
\label{subsec:Brownian_limit}
\begin{figure*}
    \centering
    \includegraphics{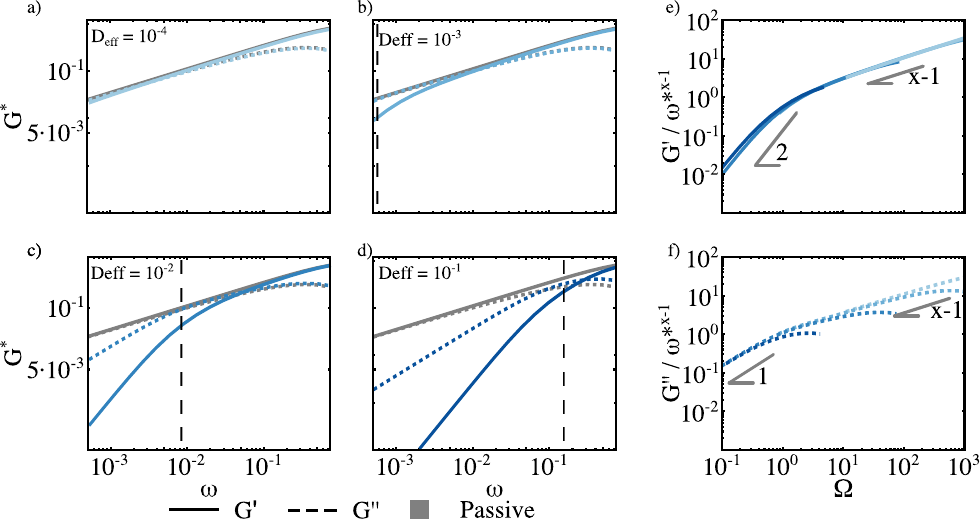}
    \caption{a-d) Complex shear modulus $G^*$ in the Brownian limit, obtained for increasing values of the effective diffusion coefficient $D_{\rm eff}$, as shown in the figures. Activity induces a crossover at some frequency $\omega^*$ discussed in the main text and indicated by the vertical black dashed lines. As in Fig.~\ref{fig:spectrum_adiabatic_limit}, at frequencies below $\omega^*$ the viscoelastic spectrum exhibits Maxwell behaviour, while it approaching passive SGR behaviour (gray lines) at higher frequencies.
    (e,f) Rescaled moduli vs rescaled frequency $\Omega = \omega / \omega^*$. The curves are color-coded as in (a-d). The data are consistent with collapse to a master curve for small $\omega^*$.}
\label{fig:spectrum_diffusive_limit}
\end{figure*}

\noindent 
For small persistence time $\tau_p$, the OU process for the active shear rate $v$, as described by eq.~(\ref{eqn:OU_activity}), reduces to Gaussian white noise of variance $2 v_0^2 \tau_p$.
The dynamics of the local strain $\ell$ between yields is therefore governed by a Langevin equation with drift $\dot\gamma$ and %approaches diffusive dynamics, with 
effective diffusion coefficient $D_{\rm eff} = v_0^2 \tau_p $.
Accordingly the active SGR model dynamics~(\ref{eqn:P_master_equation_with_activity}) can be reduced in the small $\tau_p$ limit to a master equation for $P(E,\ell;t)$ where $v$ no longer appears explicitly:
    \begin{equation}
\label{eqn:P_master_equation_with_activity_Brownian}
        \begin{aligned}
                \partial_t P(E,\ell;t) = &- \dot\gamma\partial_\ell\, P 
                +D_{\rm eff}\partial_l^2 P 
                - Y(E,\ell)P +\langle Y(E,\ell) \rangle \rho(E) \delta(\ell) %\\&
                .
        \end{aligned}
    \end{equation}
As for the adiabatic limit we then first need to determine the steady state distribution $P_{\rm ss}(E,\ell)$ in the absence of external strain, from
    \begin{equation}
    \label{eqn:diffusion_pde}
        D_{\rm eff} \partial_\ell^2 P_{\rm ss} - Y(E,\ell) P_{\rm ss} + Y_{\rm ss} \rho(E) \delta(\ell) = 0 .
    \end{equation}
This is an inhomogeneous second-order differential equation for $P_{\rm ss}(E,\ell)$. The yield energy $E$ appears only as a parameter, with the solutions for different values of $E$ eventually being coupled by the overall normalization via $Y_{\rm ss}$.
Nonetheless, because of the complicated $\ell$-dependence of the yield rate~(\ref{eqn:yield rate}), there is no straightforward analytical solution of eq.~(\ref{eqn:diffusion_pde}).

To make progress we approximate $Y(E,\ell)$ by a piecewise constant function that is in fact a lower bound:
    \begin{equation}
        Y(E,\ell) \approx 
        \begin{cases}
            e^{-E/x} \quad &\text{for\ \ \ } \ell < \sqrt{2 E}\\
            1 \quad &\text{for\ \ \ } \ell \geq \sqrt{2 E}
        \end{cases} .
        \label{eq:Y_approx}
    \end{equation}
Intuitively, this means that for strains below the local yield strain we approximate yielding as taking place only by activation via mechanical noise, with the same energy barrier as for zero strain. Once the yield strain is reached, on the other hand, the yield rate is $\Gamma_0=1$ as prescribed by eq.~(\ref{eqn:yield rate}).
With the above approximation, eq.~(\ref{eqn:diffusion_pde}) becomes a linear differential equation with coefficients that are constant in the regions $0<|\ell|<\ell_y=\sqrt{2E}$ and $|\ell|>\ell_y$. It can therefore be solved by linear combinations of exponentials, which are connected at the boundaries of the regions by the appropriate continuity requirements for $P$ and $\partial_\ell P$.

By analogy with our treatment of the adiabatic limit, once $P_{\rm ss}(E,\ell)$ has been found for the case without external strain, one can add a small oscillatory strain $\gamma(t)$ and find the function $P^{(1)}(E,\ell;\omega)$ that determines the resulting oscillatory perturbation of the distribution of $E$ and $\ell$; from this $G^*(\omega)$ can then be calculated as the average of $\ell$.
    
Fig.~\ref{fig:spectrum_diffusive_limit}(a-d) shows exemplary numerical results for $G^*(\omega)$ from the approximate analysis of the Brownian limit described above. As in the adiabatic limit, we observe a broad crossover in frequency between low-frequency Maxwell and higher-frequency SGR power law shear rheology. 
To estimate the frequency where this crossover takes place, we compare as before the relevant yielding timescales:  the one for yielding activated by mechanical noise, which is $\tau(E)$, and the one for yielding by diffusion of the local strain $\ell$, caused by the rapidly fluctuating active shear rate. This diffusive yielding process needs to increase $\ell$ from zero to $\ell_y$, so takes a time of order $\ell_y^2/(2 D_{\rm eff}) =
E/D_{\rm eff}$.
Equating therefore $\tau(E) = E/D_{\rm eff}$ to determine the location of the crossover, one finds again a solution in terms of Lambert functions, whose leading order behaviour is 
\begin{equation}
\tau^* = \frac{x}{D_{\rm eff}} \ln( x/D_{\rm eff})\ .
\label{eq:taustar_Brownian}
\end{equation}
The corresponding crossover frequency estimate $\omega^*=2\pi/\tau^*$ is shown in Fig.~\ref{fig:spectrum_diffusive_limit} by the black dashed lines, and gives a reasonable account of the numerically observed change between the two limiting power laws. As in the adiabatic limit, one then further expects that a scaling limit must exist for small $\omega^*$, and the appropriately scaled results in Fig.~\ref{fig:spectrum_diffusive_limit}(e,f) are consistent with this expectation. 

It remains, finally, to check how accurate our analysis of the Brownian limit is, given that we used the approximate form~(\ref{eq:Y_approx}) of the yield rate. To assess this, we have directly simulated the stochastic dynamics of $E$ and $\ell$ described by eq.~(\ref{eqn:P_master_equation_with_activity_Brownian}) and determined the resulting viscoelastic moduli $G^*(\omega)$ numerically. The data (not shown here) do of course exhibit some quantitative deviations from those in Fig.~\ref{fig:spectrum_diffusive_limit}, but in qualitative terms they show all the same trends.

\subsection{Activity with finite persistence}
\label{subsec:Finite_persistence}
We have, so far, analysed two limits of the active SGR model that are largely analytically tractable, owing to a timescale separation between the dynamics of the activity and that of the local strains and yield energies.
For intermediate values of the persistence time $\tau_p$ of the activity, we were unable to find even an approximate closed form solution for the viscoelastic moduli. 

However, given that both in the small and large $\tau_p$ limits we found viscoelastic spectra crossing over from Maxwell relaxation at low frequencies to SGR power laws at higher frequencies, it is natural to expect the same behaviour at intermediate $\tau_p$. The main question then becomes how the crossover timescale $\tau^*$ or the corresponding frequency $\omega^*$ depend on the activity parameters. Omitting logarithmic factors, we found above in eq.~(\ref{eq:taustar_Brownian}) for low $\tau_p$ that $\tau^* \approx x/D_{\rm eff}=x/(v_0^2\tau_p)$. For large $\tau_p$, on the other hand, the corresponding result from eq.~(\ref{eq:taustar_adiabatic}) is $\tau^*\approx \sqrt{x}/v_0$. These two limits suggest as a simple interpolation for intermediate $\tau_p$
\begin{equation}
\tau^* \approx \frac{x}{v_0^2} \frac{1 + v_0 \tau_p / \sqrt{x}}{\tau_p} .
\label{eq:taustar_interpolation}
\end{equation}
From this one sees that the dimensionless parameter determining the transition between the Brownian and adiabatic regimes is $v_0\tau_p/\sqrt{x}\sim v_0\tau_p$. This has a simple interpretation as the typical amount of active strain that can be accumulated because $v(t)$ reverses sign: if this quantity is large then large active strains can result and correspondingly one gets behaviour approaching the adiabatic limit, while for small $v_0\tau$ the activity acts essentially as noise in the time evolution of the local strain, so that the Brownian limit is approached.

We have verified by numerical simulations (data not shown) that the estimate~(\ref{eq:taustar_interpolation}) gives a good account of the linear viscoelastic properties predicted by the active SGR model for intermediate $\tau_p$. For these simulations we implement the Langevin dynamics of $v$ using the Euler-Maruyama scheme and then feed this into the dynamical evolution of $E$ and $\ell$.
The time-domain stress relaxation modulus $G(t)$ is determined by application of a small step strain, and from this the complex shear modulus is derived using the relation $G^*(\omega) = i \omega \int_0^{\infty} dt \, G(t) e^{- i \omega t}$, with the Fourier transform implemented using the Fast Fourier Transform (FFT) scheme.
    
\section{Discussion and Conclusions}
\label{sec:Discussion_Conclusions}
Inspired by the action of molecular motors on cytoskeletal filaments \cite{Ahmed_2015,Fodor_2015,Pegoraro_2017}, we introduced in this paper an active SGR model where activity acts as an independent, stochastic local strain rate. This is characterized by an amplitude $v_0$ and a persistence time $\tau_p$, and we showed that the active drive generically induces a long-time fluidisation of the soft glassy material. In the viscoelastic spectrum, this appears as a crossover from the usual SGR-like power-law response to a Maxwell-like relaxation at low frequencies. We identified this crossover analytically in two limiting regimes: the adiabatic limit of highly persistent activity, where the active strain rate is approximately constant, though still stochastic, on typical yielding times; and the Brownian limit of short persistence time, where the active drive reduces to an effective diffusivity of the local strain.
In both cases, the crossover frequency can be interpreted in terms of an activity-induced timescale
separating two populations of local elements or ``traps'': shallow traps, i.e.\ elements with low yield energy $E$, relax predominantly through the active Maxwell channel, whereas deep traps with large $E$ remain controlled by the Arrhenius-like SGR dynamics at noise temperature $x$. The crossover frequency $\omega^*\sim 1/\tau^*$ corresponds to the timescale at which these two relaxation mechanisms become comparable. Interestingly, its dependence on the activity parameters $v_0$ and $\tau_p$ is quite distinct in the Brownian and adiabatic limits: 
for short persistence time $\tau_p$ we find $\tau^* \sim x/(v_0^2\tau_p)$. The numerator reflects the integrated active-noise strength, which grows with $\tau_p$ and so leads to smaller $\tau^*$. Activity-induced fluidisation is thus observed up to increasing frequencies $\omega^*$ as $\tau_p$ grows.
In the adiabatic limit, on the other hand, we obtain $\omega^*\sim {v_0}/{\sqrt{x}},$
showing that fluidisation saturates as $\tau_p$ increases.
Interpolating between the two limits, we saw that the relevant dimensionless parameter governing the transition between them is $v_0\tau_p$: 
once the active strain accumulated over one persistence time becomes large enough ($v_0\tau_p>1$) to cause typical elements to yield, increasing $\tau_p$ does not change the physics further and the rheology approaches the adiabatic limit.

Similar rheological crossovers have recently been reported in simulations of active Brownian particles, where Martín-Roca et al.~found a Maxwell-like response with a crossover frequency controlled by activity and a nonmonotonic dependence once motility-induced phase separation sets in \cite{Roca_2025}. Our model does not contain phase separation and predicts saturation rather than nonmonotonicity. Nevertheless, both results point to the emergence of an activity-controlled structural relaxation time in dense active materials. A closer analogy can be drawn with active trap models driven by OU forces, where the effective activation timescale for escape crosses over from a Brownian regime controlled by the integrated active noise to an adiabatic regime independent of the persistence time \cite{Woillez_2020}. In our case, the same physical mechanism appears not as an effective activation temperature, but directly as an inverse fluidisation time in the rheological response. A further distinction is that in the case of active OU particles the change between Brownian and adiabatic regimes arises from the competition between the persistence time and the (passive) relaxation time in the trap. In the active SGR model, on the other hand, the competition is between two parameters of the activity itself, $1/v_0$ and $\tau_p$, which each define a timescale.

More broadly, our results connect to effective-temperature descriptions of active matter, although with important distinctions. Active mode-coupling theory and active random first order transition theory predict analogous saturation of activity-induced effective temperatures for highly persistent activity at fixed variance. In the former case, a global effective temperature is obtained from a generalized fluctuation-dissipation relation \cite{Nandi_2017,Ghosh_2025}, whereas in the latter it emerges from the activity-induced modification of the single-particle configurational entropy in a cage \cite{Nandi_2018}. Although these approaches refer to different physical observables, both predict that the effect of activity saturates at high persistence.

In the opposite limit of small persistence, activity generates a effective diffusivity, as previously noted for dilute ABP systems \cite{Bechinger_2016}. This allows one to introduce an effective temperature $T_{\rm eff}\sim v_0^2\tau_p$ through a generalized Stokes-Einstein relation, with the same scaling as our crossover frequency. Similarly, Loi et al.~found that an actively kicked Langevin system exhibits an effective increase of the bath temperature, again saturating for probes coupled to the slow degrees of freedom \cite{Loi_2008}. We emphasize, however, that in the active SGR model the central quantity is not a thermodynamic effective temperature, but an activity-induced relaxation rate. The common feature is that activity dominates the long-time dynamics: at small persistence through diffusive active noise, and at large persistence through persistent threshold crossing, leading to saturation of the fluidisation rate.

Looking ahead, our analysis of the active SGR model introduced above has been restricted to the linear rheological regime. Since biological tissues and cytoskeletal materials often experience large deformations under physiological conditions \cite{Fernandez_2017,Fielding_2023,Pegoraro_2017}, an important direction for future work is to investigate the nonlinear response, with preliminary results showing nontrivial effects arising from the capped yield rate \cite{in_preparation}. It will be an interesting question to study the interplay of these effects with activity. Extensions of the active SGR model to more elaborate forms of activity \cite{Ghosh_2025, Keta_2025} will also be useful to consider, in order to assess the robustness of the predictions from the simple model proposed here.

\section{AI documentation}
During the preparation of this work, the authors used Microsoft Copilot, Qwen 3 30B A3B Instruct  2507 and OpenAI--GPT 5/5.5 for coding support, improving readability and language phrasing. The authors have reviewed and edited the content and take full responsibility for the content of the publication.

\section*{Author contributions}
Mendozza, R.: Conceptualization, Methodology, Writing — original draft, Writing — review \& editing. Müller, T.: Data curation, Visualization. Sollich, P.: Conceptualization, Methodology, Validation, Writing — original draft, Writing — review \& editing, Supervision, Project administration.

\section*{Conflicts of interest}
There are no conflicts to declare

\section*{Acknowledgements}
We would like to thank Timo Betz for helpful discussions. This work was funded by the Deutsche Forschungsgemeinschaft (DFG, German Research Foundation) under Project-ID 449750155--RTG 2756, Project A5. 

\balance

\renewcommand\refname{References}
\newpage
\bibliography{bibliography} 
\bibliographystyle{rsc} 

\end{document}